\documentstyle[prl,aps]{revtex} \def\narrowtext{} \tighten 
\twocolumn

\begin{document}
\draft
\title{
\begin{minipage}[t]{7.0in}
\scriptsize
\begin{quote}
\raggedleft {\rm cond-mat/9712094},\\
{\rm Submitted to} {\it Phys. Rev. Lett.}
\end{quote}
\end{minipage}
\medskip
\\ A Theory of Magnets with Competing \\
Double Exchange and
Superexchange Interactions}
\author{D. I. Golosov$^{1,2}$,
 M. R. Norman$^2$,
 and K. Levin$^1$}
\address{
(1) The James Franck Institute, The University of
Chicago, 5640 S. Ellis Avenue, Chicago, IL 60637 \\
(2) Materials Science Division, Argonne National
Laboratory, 9700 S. Cass Avenue, Argonne, IL 60439}
\address{%
\begin{minipage}[t]{6.0in}
\begin{abstract}
We study the competition between ferromagnetic double exchange 
(DE) and nearest-neighbour antiferromagnetic exchange in
CMR materials. Towards this end,  a single site 
mean
field theory is proposed which emphasizes the hopping-mediated
nature of the DE contribution.  We find that the
competition between these two exchange 
interactions  leads to ferro- or
antiferromagnetic 
order with incomplete saturation of the (sub)lattice 
magnetization. 
This conclusion is in contrast 
to 
previous results in the literature which find a canted spin
arrangement under similar  circumstances.
We attribute this 
difference to  the highly anisotropic exchange interactions used 
elsewhere. 
The 
associated experimental implications are discussed. 
\typeout{polish abstract}
\end{abstract}
\pacs{PACS numbers: 75.70.Pa, 75.40.Cx, 75.30.Et, 75.10.Lp}
\end{minipage}}

\maketitle
\narrowtext
The colossal magnetoresistance (CMR) manganese oxides
have received considerable attention recently \cite{Ramirez} .  In the
CMR regime, these materials 
exhibit ferromagnetism which is generally believed to result from a 
conduction electron-mediated double
exchange (DE) mechanism\cite{Zener}.  In addition, there exists
 evidence which suggests
\cite{Aeppli,Osborn} the presence of an
antiferromagnetic  superexchange of  comparable scale.
Thus, a full understanding of   
the
magnetic order in the CMR materials  requires a treatment of 
the
competition between ferro- and antiferromagnetic interactions.  The
aim of the present paper is to address this competition and, by doing 
so, revisit earlier claims in the literature\cite{DeGennes} which
suggest that spin canting may be the most natural means of
accommodating these two opposing  interactions. 

In the literature, previous related calculations\cite{DeGennes} have 
been performed 
for a strongly
anisotropic model in which the inter- and intra-layer direct exchange
constants have  different signs \cite{undoped}.
In this paper it is assumed that direct interactions  have everywhere
the same (antiferromagnetic)  sign and  
magnitude. This is viewed as more appropriate for the ${\rm
La_{1-x}Ca_xMnO_3}$ perovskite family  
away from the x=0 endpoint, as well as for the layered manganates 
such as ${\rm La}_{2-2x}  
{\rm Sr}_{1+2x} {\rm Mn}_2 {\rm O}_7$. In view of the
considerable interest in these layered  
systems \cite{Aeppli,Osborn,Tokura},  the present calculations 
address
primarily the  two dimensional (2D) 
lattice; nevertheless, our 
2D results are qualitatively representative  of  the three dimensional
case as well. In 
the present  situation, there are no 
ferromagnetically locked layers; in this way spin fluctuations are 
enhanced, thereby leading to stronger fluctuations  in the electronic 
kinetic 
energy. Because these  fluctuations are not fully captured by  lowest
order (i.e.,  Hartree--Fock like)  treatments,  in the present paper
we  introduce a new approach to the problem.  

The present theoretical framework is based on a single site mean field 
theory 
which emphasizes the hopping-mediated nature of the DE-induced
ferromagnetic interaction.
This is to be contrasted 
with alternative approaches in the literature\cite{DeGennes,Millis1}
which implicitly  introduce a Heisenberg-like  exchange interaction to
represent these DE effects.  Our starting point is the standard
Hamiltonian  \cite{DeGennes,Anderson} derived for the case of 
infinite
Hund's rule coupling: 
\begin{eqnarray}
{\cal H}&=& - {t_0} \sum_{<i,j>}\cos \frac{\theta_{ij}}{2}\,\{c^\dagger_{i}
c_{j}+c^\dagger_j c_i\} + \frac{J_{AF}}{S^2} \sum_{<i,j>}
\vec{S}_i\cdot\vec{S}_j \nonumber \\ 
&&-\frac{H}{S} \sum_i S_i^z\,.
\label{eq:ham}
\end{eqnarray}   
Here $c_{j}$
annihilates a fermion on site $j$, $\vec{S}_i$ represents a classical
core (localized) spin ($S\gg 1$),  $J_{AF}$ is the nearest-neighbour 
antiferromagnetic
exchange integral, $H$ -- external field, and $\cos
\theta_{ij}=\vec{S}_i \cdot \vec{S_j} /S^2$. 
Throughout this paper, we use units in which the bare hopping
$t_0$, $\hbar$, $\,k_B$, $\,\mu_B$, and the lattice period 
are 
equal to unity.  In the first term of Eqn. (\ref{eq:ham}), we have
omitted Berry phase effects which are insignificant for  a single-site 
mean field treatment. 

The random distribution of localized spins leads in
Eqn. (\ref{eq:ham}) to  a highly disordered electronic hopping
problem.   In our  mean field approach we focus  
on a central  site characterized by hopping $b$ to the surrounding 
sites. This site is embedded in a medium with average hopping $t$,
 $\,\,\,t \neq b$ \cite{virtual};   for clarity  these parameters are
indicated schematically in Fig. 1. The quantities
$b$ and $t$ depend in a self consistent fashion on
the change, $\delta \Omega$, in the free energy, associated with the
change in hopping matrix   
elements $t\rightarrow b$.
This kinetic energy contribution to  $\delta\Omega$, which  can be 
evaluated 
following Ref. \cite{Lifshits}, is given by  
\begin{eqnarray}
\delta\Omega_{DE}(b,t,T)=&& \int f(\epsilon)
\xi(\epsilon) d \epsilon \label{eq:iml1}
\\
&&+ \theta(b-t) \cdot
(\varphi(z_0)-\varphi(-Dt))\,\,, \nonumber 
\end{eqnarray}
where the spectral shift function $\xi(\epsilon)$ is given by 
\begin{equation}
\pi {\rm ctg} \pi \xi (\epsilon) = - \frac{1}{\epsilon \nu(\epsilon)}
\frac{b^2}{t^2-b^2} - \frac{1}{\nu(\epsilon)} {\cal P}\int
\frac{\nu(\eta) d\eta} {\epsilon - \eta}\,, 
\label{eq:iml2}
\end{equation}
the bound
state energy $z_0<-Dt$ is the root of
\begin{equation}
1+\frac{t^2-b^2}{t^2}
\left\{-1+z\int \frac{\nu(\eta) d\eta} {z -
\eta}\right\}=0\,,
\label{eq:boundeqn}
\end{equation}
$\nu(\epsilon)$ is the density of states, 
$\,\,\varphi(z)=-T {\rm ln} \{ 1+ \exp[(\mu-z)/T]\}$,
$\mu$ is the chemical potential, and 
 $\,\,f(z)=\{\exp[(z-\mu)/T] +1\}^{-1}\,$.
Eqns.  (\ref{eq:iml1}--\ref{eq:boundeqn}) are valid for a simple 
lattice in any dimensionality D from 1 to 3;
the energy integrations are performed over
the entire conduction band, extending from $-Dt$ to $Dt$.  It should 
be stressed, that it is because of the locality of the 
perturbation (which represents a lattice analogue of an $s$-wave
scattering problem) that the quantity  $\delta\Omega$ can be 
evaluated 
{\em exactly}\cite{Lifshits}. 

In the ferromagnetic phase at $T>0$, the net energy cost of a
single-spin fluctuation is (in $2D$) given by 
\begin{eqnarray}
\delta \Omega_1=&&\delta \Omega_{DE}(b_1,t,T) + 4 J_{AF} \langle 
\cos
\theta_{12} \rangle_2 - H \cos \alpha_1\,- \nonumber \\
&&- 4 J_{AF} \langle \cos
\theta_{12} \rangle_{12}+ H \langle \cos \alpha_1 \rangle_1\,.
\label{eq:domega1fm}
\end{eqnarray}
Here, $\theta_{12}$ is the angle formed by the fluctuating spin  
$\vec{S_1}$ 
with a neighbouring spin  $\vec{S_2}$, and  $\alpha_1$ is the angle
between  $\vec{S_1}$ and the average  magnetization, $\vec{M}$ 
(see
Fig. \ref{fig:canted}). We use the notation $\langle
... \rangle_l$   
to represent 
averaging over the Boltzmann probability distribution of spin
$\vec{S_l}$,
$w_l \propto \exp(-\delta \Omega_l/T)$.  It follows that $\langle 
\cos
\theta_{12} \rangle_2=M \cos \alpha_1$, and 
\begin{eqnarray}
b^2_1 &\equiv&  \langle \cos^2 (\theta_{12}/2) \rangle_2=
(1+M \cos \alpha_{1})/2\,,\,\,\,\,\,\, \nonumber \\
t^2 &\equiv& \langle b^2_1 \rangle_1\,=(1+ M^2)/2\,.
\label{eq:btfm}
\end{eqnarray}
The central mean field equation of our formalism is given by $M = 
\langle\cos 
\alpha_1\rangle_1$.

In the ferro- and antiferromagnetic phases, it is useful to construct a 
reference framework with which to compare our results. We define 
$J_{eff}(M)$ which represents an effective $M$-
dependent exchange constant for a  Heisenberg-like magnet. The 
appropriate exchange constant can be deduced by considering small 
spin fluctuations ($|\cos \alpha - M| \ll 1$), which 
correspond to small fluctuations in the hopping matrix elements ( $|t-
b|\ll t$). A perturbation expansion 
of Eqn. (\ref{eq:iml1}), then leads to 
\begin{equation}
\delta \Omega_{DE} (b,t,T) \approx -2 \frac{t-b}{t} \int
\epsilon f(\epsilon) \nu(\epsilon) d \epsilon = 2 (t-b) | E_0 |\,, 
\label{eq:smallper}
\end{equation}
 at leading order in $T/t$, where $E_0$ is the kinetic energy of the
carriers for $t=1$. 
In the ferromagnetic state, it follows from Eqn. (\ref{eq:smallper}) 
that 
\begin{equation}
J_{eff}(M)=J_{AF}-
\frac{1}{8}|E_0| \cdot
\sqrt{\frac{2}{1+M^2}}\,. 
\label{eq:JDET}
\end{equation}

The second term in the above equation represents the  DE
contribution. This term, which is  contained in other
mean field  
schemes 
\cite{Millis1,Sarker},  increases as $M$
decreases. As a consequence,  for  moderately strong 
antiferromagnetic
exchange interactions 
$J_{eff}(M)$ changes sign
as $M$ varies from $0$ to $1$. This behaviour has important
consequences: it leads to a lack of saturation in the low temperature
magnetization. Typical results for $M(T)$ are plotted in
Fig. \ref{fig:DEFM} for these moderately strong exchange interactions 
($|E_0| + 2 H < 8J_{AF} < \sqrt{2} |E_0| + 2 H$). Here 
the solid line represents the full calculation, while  the dashed line
is obtained using the effective exchange interaction.  For  
comparison we plot (dotted line) the magnetization of a 
conventional Heisenberg magnet with the same $T_C$.

The lack of saturation seen in
Fig. \ref{fig:DEFM} can be understood as follows. As $T$ decreases, 
the 
magnitude of single spin fluctuations also
decreases. This leads to an  increase of $M$, which in turn implies a 
decrease in
$|J_{eff}|$ (and thereby a tendency to {\em decrease} $M$).  Thus, 
through 
this self-adjustment of the effective 
exchange interaction (which never becomes large in comparison with
$T$), the magnetization fails to reach its proper saturation value,
$M_0=1$ \cite{M0}.  
These self consistent changes in $|J_{eff}|$ lead to inadequacies of the 
effective exchange approximation at low $T$. As may be seen in  Fig. 
\ref{fig:DEFM},  the behaviour obtained in this approximation  differs 
significantly from that found  using the full calculation of $M(T)$.  
This difference is due to the fact that 
when $J_{eff}
\stackrel {<}{\sim} T$ is small,   quadratic terms  (in $(t-b)/t$) 
dominate the physics \cite{LowT}.

The N\'{e}el antiferromagnetic state (of the metallic phase) can be 
treated 
similarly \cite{antiferro}. It can 
be shown that  N\'{e}el ordering (which 
arises  for 
$J_{AF}> 
2^{-5/2} |E_0|$ in zero field),  always exhibits undersaturation  of the 
sublattice magnetization. This undersaturation (which leads to a 
finite bandwidth) may be 
viewed as
consistent with the presumed  metallic state.

Our discussion thus far has not included the canted phase first 
proposed by 
De Gennes\cite{DeGennes}. 
This phase is characterized by spin ordering 
with two equal sub-lattice   
magnetizations $m$ which  form  an angle $2 \gamma$ between 
them.  In the present model, spin
canting  requires the presence of a  magnetic
field to break the high degeneracy which would otherwise occur.  
This degeneracy is related to the fact that  the energy of
the system depends solely on 
the cosine of the angle which the spin $\vec{S}_1$ (of sublattice I)
forms with its  nearest
neighbours belonging to sublattice II.  In the context of single site 
mean field approaches, this energy does not change as the spin
$\vec{S}_1$ moves along  any cone around the average direction of 
spins of 
sublattice II. Thus, on  average  the spin
$\vec{S}_1$ will be aligned with sublattice II, rather than
I. Therefore, in the absence of perturbations
(caused by next-nearest-neighbour exchange, anisotropy effects,
quantum corrections, or small  
external fields) the canted state is destabilized.   Since the
underlying degeneracy is
site-local \cite{Kagome}, its effects  will be suppressed  only when the energy
scale of a perturbation {\em per individual spin} is comparable with
that of the 
thermal motion of a {\em single} spin, that is, with the
temperature $T$. 

To characterize the finite field canted state, we use 
the full non-perturbative expression (\ref{eq:iml1}), with
appropriate modifications to Eqns. (\ref{eq:domega1fm}--
\ref{eq:btfm})
\cite{modification}.
We obtain two coupled self consistent equations, one for the 
component
of $\langle \vec{S}_1 \rangle$ parallel to the magnetization of
sublattice I,  
\begin{equation}
-\sin 2 \gamma \,\langle \sin \alpha_1 \, \cos \beta_1 \rangle_1 + 
\cos 2
\gamma \,\langle \cos \alpha_1 \rangle_1=m\,, 
\label{eq:mfec1} 
\end{equation}
and another for the perpendicular component,
\begin{equation}
\cos 2 \gamma \, \langle \sin \alpha_1 \, \cos \beta_1 \rangle_1 + 
\sin 2
\gamma \, \langle \cos \alpha_1 \rangle_1 =0\,.
\label{eq:mfec2}
\end{equation}
where $\alpha_1$ and $\beta_1$ are polar and azimuthal angles of 
the
spin $\vec{S_1}$ measured with respect to the co-ordinate system 
that has, as its polar axis, the 
average
direction of the spins of sublattice II.  We choose  $\beta_1=0$ for
the spin $\vec{S_1}$ lying within the plane 
containing the two sublattice magnetizations.  

The low-$T$ canted state is found to be stable for $8 
J_{AF}>|E_0| +  H$. 
The  solutions of Eqns. (\ref{eq:mfec1}--\ref{eq:mfec2}) for typical
parameters   are 
illustrated  in the inset of
Fig. \ref{fig:DEFM}.  One can see  that,
as $T \rightarrow 0$ in the canted phase, the sublattice 
magnetization $m$
approaches its proper saturation value $m=1$.  Note that the
ferromagnetic  ($\gamma =0$) 
solution to the mean field equations is
present at $H>0$ as well. In Fig. \ref{fig:DEFM} (inset), the
corresponding magnetization, $M_{FM}(T)$, is represented by the 
dotted line.  
However, when the canted ($\gamma > 0$) solution exists, it 
corresponds
to a lower value of the free energy.  The canted solution branches 
from the
ferromagnetic one at  
temperature  $T
\sim H $; at this point the
undersaturated ferromagnetic state undergoes a second-order spin-flip
transition into the low temperature canted state \cite{AFM}.
This lends support to the notion that undersaturation is
characteristic of the generic low temperature regime 
\cite{Chubukov}.

Typical phase diagrams for the DE--superexchange magnet in (left) zero 
and (right) non-zero field  
are presented  in
Fig. \ref{fig:phase}. For $t_0$ of the order of an eV, our
choice of parameters corresponds to reasonable values of $J_{AF}
\stackrel{<}{\sim}300 {\rm K}$. In zero field (left),
the solid line represents the phase boundary
between paramagnetic
(PM) and antiferro- (AFM) or ferromagnetic (FM) metallic phases.
The ordered phases are undersaturated at low $T$ (for slightly smaller
$J_{AF}$ we find a 
critical value of bandfilling,  $x_1$, which divides the saturated,
$x>x_1$, and undersaturated regimes).
At low temperatures and small concentrations (in 2D, $x < 0.215$),
the undersaturated AFM state becomes thermodynamically
unstable ($\partial \mu /\partial x < 0$),  signalling either the
onset of a more
complicated spin arrangement or phase separation. The dashed line in
Fig. \ref{fig:phase} corresponds to the boundary of this region
($\partial \mu /\partial x = 0$).

The right panel in Fig. \ref{fig:phase} shows that in the
presence of a magnetic field the PM--FM transition is replaced by a
smooth crossover (dotted line).  The spin arrangement  
of the AFM phase becomes non-collinear (flop-phase), and has
the same symmetry properties as the canted phase (CM), 
which becomes stable at lower $T$ (replacing the $H=0$ 
undersaturated FM and AFM phases). The two are separated from the PM
and FM region by a second-order phase transition, which is represented
by the solid line. At
sufficiently small $x$ the latter approaches the $H=0$  N\'{e}el
transition line. The thermodynamic instability line (not shown)
is only slightly affected by $H$.

These calculations have made a number of predictions which can be
tested experimentally. The layered materials ${\rm La}_{2-2x}  
{\rm Sr}_{1+2x} {\rm Mn}_2 {\rm O}_7$,   with $x=0.4$,  presumably 
lie  either
within the region where the system should display
undersaturated ferromagnetic behaviour at low $T$, or on the brink 
of this
region, where  thermal fluctuations should still be stronger than 
in a
conventional magnet. 
Some measurements of the absolute value of magnetization in  $x=0.4$
samples support undersaturation \cite{Rietveld,Raveau}, while others
do not \cite{Moritomo}. 
We also note 
that the presence of undersaturation in ferro- and antiferromagnetic
phases may well signal that in reality the system favours more
complicated ({\em e.g.} spin glass-like, cf. Ref. \cite{Moritomo})
spin ordering, that cannot be addressed within a single-site mean
field theory.  


It is natural to expect that the relative strength
of the superexchange interaction is even higher in $x=0.3$ 
compounds,
so that this compound may be suitable for observing canting under 
the
proper field and temperature conditions:  $T \stackrel {<}{\sim}H$,
although experiments have not yet been performed in this regime.
There have been no observations of  
an {\em ordered} canted state (as distinguished from possible canting
correlations reported in Ref. \cite{Osborn})  in the layered compounds.
This is consistent with our results.  
Finally, we propose that the
magnetization dependence of the effective exchange constant 
(available
through spin wave measurements) should be studied in both 3D and 
2D systems. 

This work has benefited from enlightening discussions that we had
with many theorists and experimentalists, especially  A. G. Abanov,
A. Auerbach, 
A. V. Chubukov, M. I. Kaganov, R. Osborn, R. M. Osgood, and A. E. 
Ruckenstein. We
acknowledge the support of a Univ. of Chicago/Argonne National
Lab. collaborative 
Grant, U. S. DOE, Basic Energy Sciences, Contract
No. W-31-109-ENG-38, and the MRSEC program of the NSF under 
award \#
DMR 9400379.

\begin{figure}
\caption{Single-spin fluctuation in the ferromagnetic phase. The bold 
arrow
represents the average magnetization, and the dashed lines correspond 
to the
hopping amplitude $b$, which differs from the background hopping 
value 
$t$
(solid lines).}
\label{fig:canted}
\end{figure}

\begin{figure}
\caption{Magnetization vs. temperature in the ferromagnetic phase 
at $H=0$,
$x=0.4$, and  $J_{AF}=0.06$. The
solid, dashed, and dotted lines correspond to the 2D DE--
superexchange
magnet, effective exchange approximation, and usual Heisenberg
ferromagnet, respectively. The inset
shows the behaviour of the sublattice (solid line) and net (dashed line,
$M=m\cos \gamma$)
magnetizations in the canted state at $H=0.01 $ in comparison with 
the magnetization of the ferromagnetic state (dotted line).} 
\label{fig:DEFM}
\end{figure}

\begin{figure}
\caption{Phase diagrams of the DE--superexchange magnet for
$J_{AF}=0.06$  at $H=0$ (left panel) and $H=0.01$ (right panel),
showing the ferro-, antiferro- (flop-phase at $H>0$), paramagnetic,
and canted phases (FM, AFM, PM, and CM, respectively). The
behaviour of the system is symmetric with respect to quarter-filling, 
$x=0.5$.}
\label{fig:phase}
\end{figure}


\begin{references}
\bibitem{undoped} This model was chosen to fit the known magnetic 
behaviour of the 
undoped ${\rm La Mn O}_3$.
\bibitem{virtual} We are, in effect, using the  virtual
crystal approximation,   
which is expected to be appropriate to a description of the
metallic phases, as long as the carrier concentration is not
too small. We note that our choice of parametrization would have to be
generalized to incorporate a
$(\pi,0,0)$-type AF ordering.
\bibitem{M0} Instead, the limiting  value of the zero field, zero
temperature magnetization  
is given by
$M_0=\{(E_0/J_{AF})^2/32-1\}^{1/2} < 1$.
\bibitem{antiferro} In this case,
$\langle \cos \theta_{12} \rangle_2 = - m \cos \alpha_1$, where $m$ is
sublattice magnetization and $\alpha_1$ is the angle formed by
the spin $\vec{S}_1$ with its average direction.  Eqns.
(\ref{eq:btfm}) are replaced by $b_1^2=(1-m \cos \alpha_1)/2$ and 
$t=(1-m^2)/2$. Instead of (\ref{eq:JDET}) we obtain
$J_{eff}(m)=J_{AF}-|E_0|\{2/(1-m^2)\}^{1/2}/8$.
\bibitem{Kagome} Another example of a classical spin system with
site-local continuous degeneracy is provided by the Kagom\'{e} 
antiferromagnet.
\bibitem{modification} Instead of
Eqns. (\ref{eq:domega1fm}--\ref{eq:btfm}), we obtain $\delta 
\Omega_1=\delta 
\Omega_{DE}(b_1,t,T) + 4 J_{AF} m \cos \alpha_1- H(-\sin 
\gamma\,\sin
\alpha_1\,\cos \beta_1+\cos\gamma \cos \alpha_1)$, 
$\,\,b_1^2=(1+m\cos
\alpha_1)/2$, and $t^2=(1+m^2 \cos 2 \gamma)/2\,$.
\bibitem{Chubukov} While it is not
clear whether  undersaturation is present in the case of quantum
spins, we note that
this would be not inconsistent with the results obtained for the somewhat
similar case of quantum ferrimagnets (see: A. V. Chubukov {\em et
al.}, J. Phys.: 
Cond. Matt. {\bf 3}, 2665 (1991)).
\end{references}
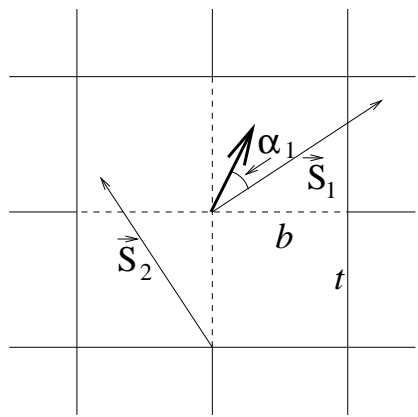
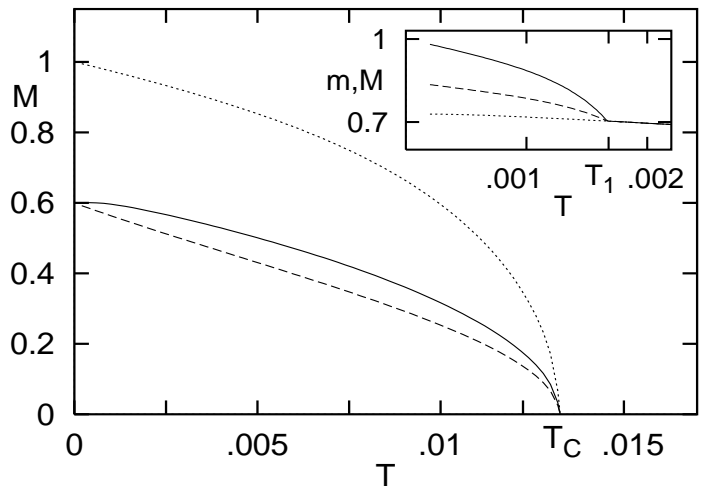
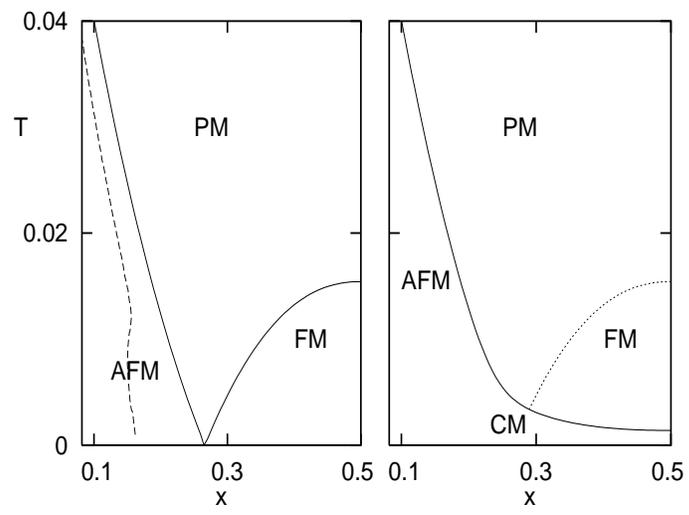
\end{document}